\documentclass[apj]{emulateapj}

\usepackage{color}
\usepackage{natbib}
\usepackage{color}
\usepackage[colorlinks, allcolors=blue]{hyperref}

\begin{document}

\title{IRIS Burst Spectra Co-Spatial To A Quiet-Sun Ellerman-Like Brightening}
\author{C. J. Nelson\altaffilmark{1,2}, N. Freij\altaffilmark{3}, A. Reid\altaffilmark{2}, R. Oliver\altaffilmark{3,4}, M. Mathioudakis\altaffilmark{2}, R. Erd\'elyi\altaffilmark{1,5}.}

\email{c.j.nelson@sheffield.ac.uk}
\shorttitle{IRIS bursts co-spatial to QSEBs}
\shortauthors{Nelson \it{et al.}} 

\altaffiltext{1} {Solar Physics and Space Plasma Research Centre, University of Sheffield, Hicks Building, Hounsfield Road, Sheffield, UK, S3 7RH.}
\altaffiltext{2} {Astrophysics Research Centre (ARC), School of Mathematics and Physics, Queen’s University, Belfast, BT7 1NN, Northern Ireland, UK.}
\altaffiltext{3} {Departament de F\'isica, Universitat de les Illes Balears, 07122 Palma de Mallorca, Spain.}
\altaffiltext{4} {Institut d'Aplicacions Computacionals de Codi Comunitari (IAC$^3$), Universitat de les Illes Balears, 07122 Palma de Mallorca, Spain.}
\altaffiltext{5} {Department of Astronomy, E\"otv\"os Lor\'and University, Budapest P.O. Box 32, H-1518, Hungary.}

\begin{abstract}
Ellerman bombs (EBs) have been widely studied over the past two decades; however, only recently have counterparts of these events been observed in the quiet-Sun. The aim of this article is to further understand small-scale quiet-Sun Ellerman-like brightenings (QSEBs) through research into their spectral signatures, including investigating whether the hot signatures associated with some EBs are also visible co-spatial to any QSEBs. We combine H$\alpha$ and \ion{Ca}{2} $8542$ \AA\ line scans at the solar limb with spectral and imaging data sampled by the Interface Region Imaging Spectrograph (IRIS). Twenty one QSEBs were identified with average lifetimes, lengths, and widths measured to be around $120$ s, $0.63$\arcsec, and $0.35$\arcsec, respectively. Three of these QSEBs displayed clear repetitive flaring through their lifetimes, comparable to the behaviour of EBs in Active Regions (ARs). Two QSEBs in this sample occurred co-spatial with increased emission in SDO/AIA $1600$ \AA\ and IRIS slit-jaw imager $1400$ \AA\ data, however, these intensity increases were smaller than reported co-spatial to EBs. One QSEB was also sampled by the IRIS slit during its lifetime, displaying increases in intensity in the \ion{Si}{4} $1393$ \AA\ and \ion{Si}{4} $1403$ \AA\ cores as well as the \ion{C}{2} and \ion{Mg}{2} line wings, analogous to IRIS bursts (IBs).  Using RADYN simulations, we are unable to reproduce the observed QSEB H$\alpha$ and \ion{Ca}{2} $8542$ \AA\ line profiles leaving the question of the temperature stratification of QSEBs open. Our results imply that some QSEBs could be heated to Transition Region temperatures, suggesting that IB profiles should be observed throughout the quiet-Sun.
\end{abstract}


\section{Introduction}
	\label{Introduction}

Ellerman bombs (hereafter referred to as EBs) are small-scale (lengths often below $1$\arcsec), short-lived (lifetimes below $10$ minutes) events which were originally identified by \citet{Ellerman17} as regions of intense brightness in the wings of the H$\alpha$ line profile. These features were named `petit points' by \citet{Lyot44} and `moustaches' by \citet{Severny56}, before the term `EBs' was coined by \citet{Mcmath60}. EBs have been widely observed co-spatial to regions of opposite polarity magnetic field (see, for example, \citealt{Pariat04,Watanabe11,Reid16}) and have been interpreted as the signatures of magnetic reconnection in the photosphere ({\it e.g.}, \citealt{Watanabe08,Archontis09,Yang16}). Until recently, these events had been exclusively observed within Active Regions (ARs), however, new research by \citet{Rouppe16} has indicated the presence of EB-like events in the quiet-Sun, named quiet-Sun Ellerman-like brightenings (QSEBs) to distinguish these events from EBs themselves.

\begin{figure*}
\includegraphics[scale=0.47,trim={0.5cm 0 0 0}]{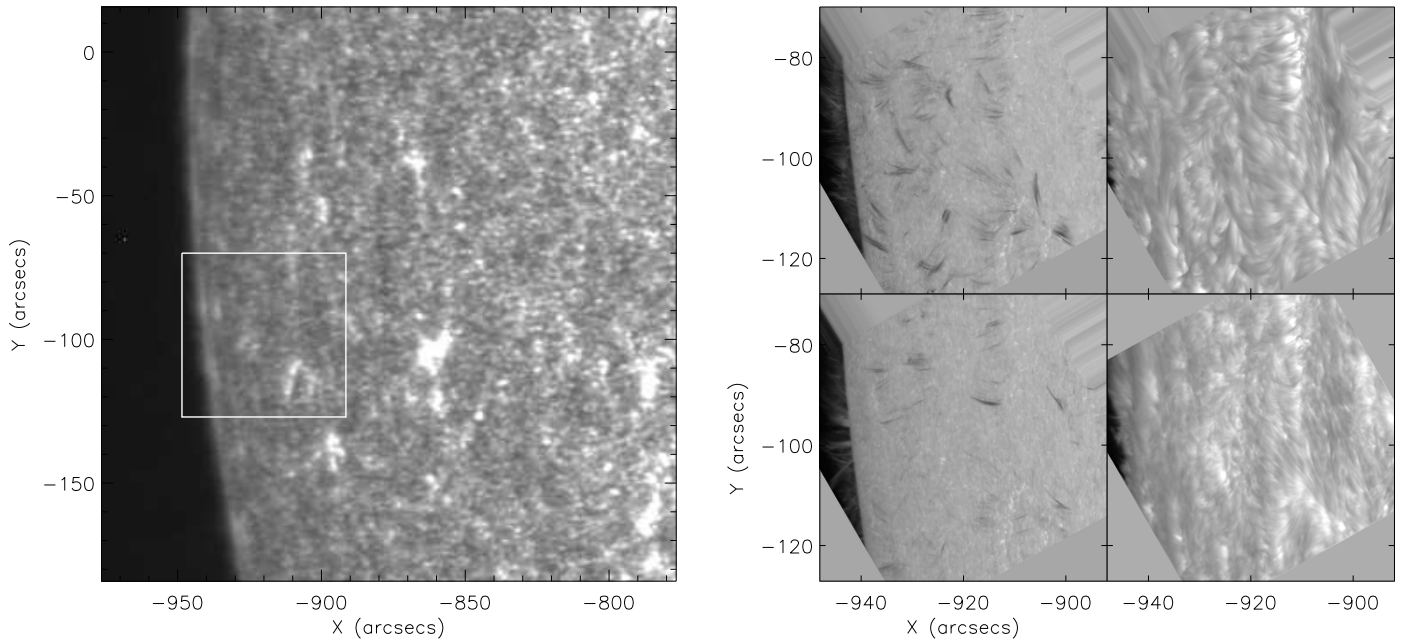}
\caption{(Left) Context SDO/AIA $1700$ \AA\ image of the region surrounding the SST/CRISP FOV (white box) sampled at $07$:$39$:$17$ UT. (Right) Clockwise from the top left are the initial (therefore, temporally closest to the SDO frame) SST/CRISP images in the: H$\alpha$ blue wing ($-1$ \AA), H$\alpha$ line core, \ion{Ca}{2} $8542$ \AA\ line core, and H$\alpha$ red wing ($+1$ \AA).}
\label{Context}
\end{figure*}

One of the most puzzling aspects of EBs is their appearance, or lack of, in a range of spectral lines. Identified signatures of these events include increased wing emission in the \ion{Ca}{2} $8542$ \AA\ (\citealt{Socas06}), \ion{He}{1} D3, and \ion{He}{1} $10830$ \AA\ (\citealt{Libbrecht16}) profiles, as well as enhanced emission in the $1600$ \AA\ (\citealt{Qiu00}) and $1700$ \AA\ continua (\citealt{Vissers13,Vissers15}). Both \citet{Ellerman17} and \citet{Rutten15}, however, identified no signatures of EBs in the \ion{Na}{1} or \ion{Mg}{1} b2 lines. Semi-empirical modelling of a variety of combinations of these lines originally led to estimates of heating within the local photospheric plasma ranging from $\sim400$-$2000$ K (see, {\it e.g.}, \citealt{Kitai83,Fang06,Berlicki14,Grubecka16}). Temperature increases of this order were challenged by results obtained through analysis of Interface Region Imaging Spectrograph (IRIS; \citealt{dePontieu14}) data which implied significantly more heating is occurring during the lifetimes of EBs (up to $8\times10^4$ K).  

The identification of small-scale brightening events in the IRIS \ion{Si}{4} `Transition Region' (TR) lines was accomplished by \citet{Peter14} who observed `hot explosions', with estimated temperatures of $8\times10^4$ K. These authors suggested that such IRIS bursts (IBs) could be evidence of heating within photospheric EBs, which could not be directly identified in that work, to temperatures an order of magnitude higher than those predicted by previous semi-empirical modelling. It should be noted, however, that such modelling had, until that point, been conducted under the assumption that EBs occurred in relatively cool photospheric and chromospheric conditions and had not, therefore, attempted to account for the high temperatures of IBs. In an independent analysis, however, \citet{Judge15} asserted that these IRIS features were, instead, formed in the chromosphere or above. Links between at least a sub-set of EBs and IBs were established by \citet{Vissers15}, \citet{Kim15}, and \citet{Tian16} appearing to support the assertion that it is photospheric plasma contained within EBs that reaches TR temperatures. It should be noted, though, that recent work by \citet{Rutten16}, who assumed that the visibility of EBs could be explained by LTE modelling, has suggested that temperatures as low as $2\times10^4$ K could account for the observed increased emission in the \ion{Si}{4} line. 

With relation to QSEBs, \citet{Rouppe16} found no evidence of emission signatures in \ion{Ca}{2} $8542$ \AA\ data, \ion{Si}{4} images (sampled by IRIS), or the $1600$ \AA\ and $1700$ \AA\ continuum, indicating that these events could be formed at lower temperatures or physical heights than their AR cousins. The observed lengths and lifetimes of these events were also smaller than those found for AR EBs in the literature (see, for example, \citealt{Watanabe11,Vissers13,Nelson15}) leading these authors to suggest that QSEBs were a `weaker member' of the small-scale, reconnection driven family of events in the lower solar atmosphere, possibly consistent with the modelling attempts of \citet{Nelson13}. Notably though, \citet{Rouppe16} only studied slit-jaw images from IRIS, and a detailed analysis of the spectra, in order to identify whether typical IB profiles can be observed co-spatial to QSEBs, is still required.

Recently, \citet{Reid17} used RADYN simulations (\citealt{Carlsson92, Carlsson95}) to model one-dimensional solar atmospheres perturbed by energy deposition at multiple layers. These authors then synthesised H$\alpha$ and \ion{Ca}{2} $8542$ \AA\ line profiles finding that impulsive energy releases in the upper photosphere could account for EB signatures. Whether such techniques and models could reproduce QSEB signatures ({\it i.e.}, H$\alpha$ wing emission with no co-temporal \ion{Ca}{2} $8542$ \AA\ response), however, is still unknown and will be discussed here.

\begin{figure*}
\includegraphics[scale=0.49,trim={0 0 0 0}]{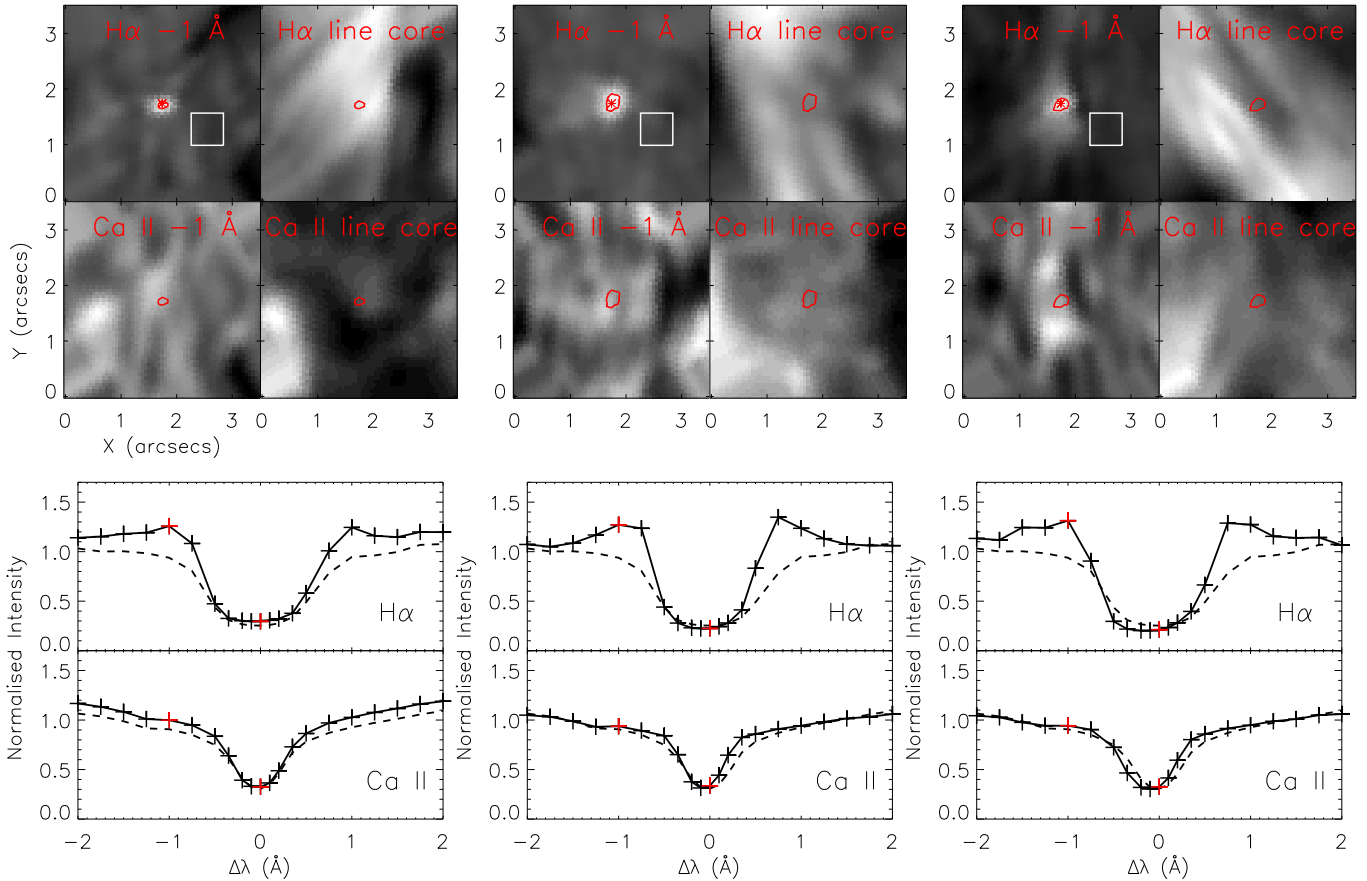}
\caption{Three representative QSEBs from the sample analysed here. (Top row) Images collected at various positions within the H$\alpha$ and \ion{Ca}{2} $8542$ line profiles (specific line positions are indicated in individual panels). The red cross and white box in the top left panel of each set indicates the pixel and pixels selected to construct line profiles of the QSEBs and (time-averaged) background atmosphere, respectively. The red contours indicate the pixels over $130$ \% of the background intensity in the H$\alpha$ line wings. (Bottom row) The H$\alpha$ and \ion{Ca}{2} $8542$ \AA\ line profiles measured for the QSEB and the background region. Red crosses indicate the wavelength positions plotted in the top row.}
\label{Profiles}
\end{figure*}

In this article, we aim to further understand both QSEBs and IBs, specifically by researching whether IBs are also evident in the quiet-Sun co-spatial to any QSEBs. We structure our work as follows: In Section~\ref{Observations}, we introduce the observations analysed here. Section~\ref{Results} presents our results, including the inference of the basic properties of QSEBs and an analysis of the signatures of one of these events in spectra collected by IRIS. Finally, we draw our conclusions in Section~\ref{Discussion}.

\section{Observations}
	\label{Observations}

The ground-based observations analysed in this article were acquired with the {\it CRisp Imaging SpectroPolarimeter} (CRISP: \citealt{Scharmer06,Scharmer08}) at the Swedish $1$-m Solar Telescope (SST: \citealt{Scharmer03}) on $9$th June $2016$. A quiet-Sun region (co-ordinates of $x_\mathrm{c}\approx-900$\arcsec, $y_\mathrm{c}\approx-100$\arcsec) was selected for observation between $07$:$39$:$29$ UT and $08$:$28$:$54$ UT. The observing sequence applied during this time consisted of a $21$ point H$\alpha$ line-scan and a $21$-point full-Stokes \ion{Ca}{2} $8542$ \AA\ line scan, both of which sampled $\pm2$ \AA\ into the wings of the lines. Wideband images were also acquired for H$\alpha$ and \ion{Ca}{2} $8542$ \AA\ at each time-step for alignment purposes. These data were reduced using the multi-object multi-frame blind deconvolution (MOMFBD; \citealt{Noort05}) method, with eight exposures at each wavelength position, and following the CRISPRED pipeline (see \citealt{delaCruzRodriguez15}). Analysis was conducted, in part, using the CRISPEX package (\citealt{Vissers12}). The final science-ready cadence and pixel-scale of these data were $26.5$ seconds and $0.058$\arcsec, respectively.

Co-spatial and co-temporal data from three filters ($304$ \AA, $1600$ \AA, and $1700$ \AA) of the Solar Dynamics Observatory's {\it Atmospheric Imaging Assembly} (SDO/AIA; \citealt{Lemen12}) instrument were also analysed. A $100$\arcsec$\times100$\arcsec\ region centred on the CRISP/SST field-of-view (FOV) was downloaded for the entire time-series. The pixel-scale of these data is approximately $0.6$\arcsec\ (corresponding to around $435$ km in the horizontal scale). The cadence of these data is wavelength-dependent, being $12$ seconds for the $304$ \AA\ data and $24$ seconds for the $1600$ \AA\ and $1700$ \AA\ data. Alignment of the SDO/AIA filters with the SST/CRISP line scans was achieved by matching the solar limb and stable network bright points within the $1600$ \AA\ and wideband \ion{Ca}{2} $8542$ \AA\ context images through time. The initial FOV of these data is plotted for reference in Fig.~\ref{Context} including a larger SDO/AIA $1700$ \AA\ context image (left panel) and (clockwise from the top left in the right panel) SST/CRISP images for the H$\alpha$ blue wing ($-1$ \AA), H$\alpha$ line core, \ion{Ca}{2} $8542$ line core, and H$\alpha$ red wing ($+1$ \AA). The white box in the left panel indicates the SST/CRISP FOV.

Finally, data from the IRIS satellite were also analysed. IRIS collected five dense $320$-step rasters between $07$:$31$:$21$ UT and $11$:$39$:$14$ UT, with the first raster coinciding temporally with the SST/CRISP observations. However, as IRIS began the scan off the solar disk and then progressed across the SST/CRISP FOV, only those data sampled between $08$:$10$:$00$ UT and $08$:$20$:$00$ UT are co-temporal and co-spatial to the SST/CRISP dataset. Slit-jaw images (SJIs) were sampled by the $1400$ \AA, $2796$ \AA, and $2832$ \AA\ filters with cadences of $18.6$ s, $18.6$ s (with every fifth frame skipped to collect $2832$ \AA\ images), and $93$ s, respectively. The spatial resolution of these data was $\sim0.33$\arcsec. Alignment of these data to the SST/CRISP, and hence the SDO/AIA, FOV was completed by correlating the $2832$ \AA\ channel to wide-band \ion{Ca}{2} $8542$ \AA\ images. The IRIS spectral data had an exposure time of $\sim8$ s and a spectral dispersion of approximately $0.026$ \AA\ for both the NUV and FUV windows.

\subsection{Feature Identification}

Candidate QSEBs were selected by locating small-scale ($<2$\arcsec), short-lived ($1<$ lifetime $<15$ min) regions of intense brightness (over $130$ \% of the local background intensity) in the near wings ($\pm1$ \AA) of the H$\alpha$ line profile. This threshold is lower compared to the $150$ \% threshold used by, for example, \citet{Vissers13,Nelson15}) and was selected due to the inherent lower line wing intensity enhancements identified co-spatial to QSEBs by \citet{Rouppe16}. Next, we specified that the proportional increase in intensity from the background decreased further out in the wings ($\pm2$ \AA) in order to differentiate QSEBs from magnetic concentrations (MCs or `pseudo-EBs'; see \citealt{Rutten13}) which are known to increase the continuum intensity. Once the candidate QSEBs had been detected, the H$\alpha$ line core images were examined in order to remove features which corresponded to H$\alpha$ micro-flares (features with obvious emission in the H$\alpha$ line core). Finally, apparent explosive behaviour (rapid morphological evolutions widely associated with EBs; \citealt{Nelson15,Vissers15}) was required, allowing us to confidently remove any remaining MCs. Overall, $21$ QSEBs were located in this dataset for further study. 

Three of the QSEBs selected for analysis here are presented in Fig.~\ref{Profiles} for reference. The frames in the top row plot H$\alpha$ blue wing ($-1$ \AA), line core, \ion{Ca}{2} $8542$ \AA\ blue wing ($-1$ \AA), and \ion{Ca}{2} $8542$ \AA\ line core images co-spatial and co-temporal to individual QSEBs. A long, thin brightening reminiscent of the events studied by \citet{Rouppe16} can be observed in the H$\alpha$ line wing for each example, however, no unambiguous increase in intensity can be observed in any other panel of these plots. Slight \ion{Ca}{2} $8542$ \AA\ wing enhancement similar to one case presented by \citet{Rouppe16} were evident in two QSEBs from our sample; however, inspection of the imaging data revealed blurred patches, dissimilar to the compact H$\alpha$ features in both time and space, implying that these co-spatial brightenings may not be linked to the QSEB itself. The red crosses and white boxes in the top left panels of each column indicate the pixel and set of pixels used to construct QSEB (solid line) and time-averaged quiet-Sun reference (dashed line) profiles, respectively. These profiles are plotted in the bottom panels of Fig.~\ref{Profiles}, where red crosses indicate the wavelength positions plotted in the upper row. Obvious increases in intensity are evident in the wings of the H$\alpha$ line profile (peaking at approximately $\pm1$ \AA) for each of these events confirming their QSEB-like nature. The event plotted in the right-hand column is the feature sampled by the IRIS slit during its lifetime.

\section{Results}
	\label{Results}

\subsection{Properties of QSEBs derived from imaging data}

\begin{figure}
\includegraphics[scale=0.5,trim={1cm 0 0 0}]{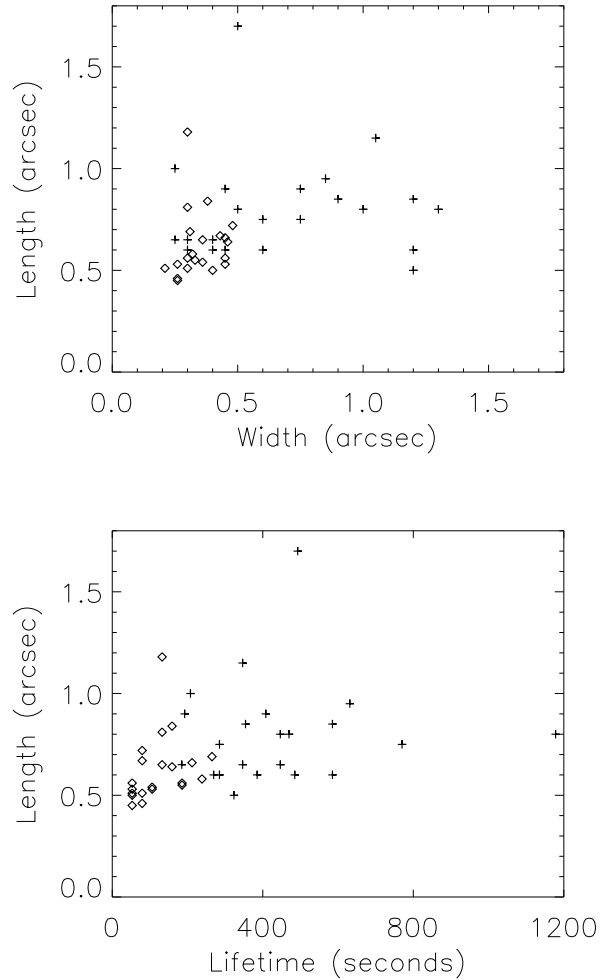}
\caption{(Top row) Length against width plot for the QSEBs discussed in this article [diamonds] and EBs analysed by \citet{Nelson15} [crosses]. (Bottom row) Same as for the top row except for length against lifetime.}
\label{Stats}
\end{figure}

\begin{figure*}
\includegraphics[scale=0.54,trim={1.5cm 0 0 0}]{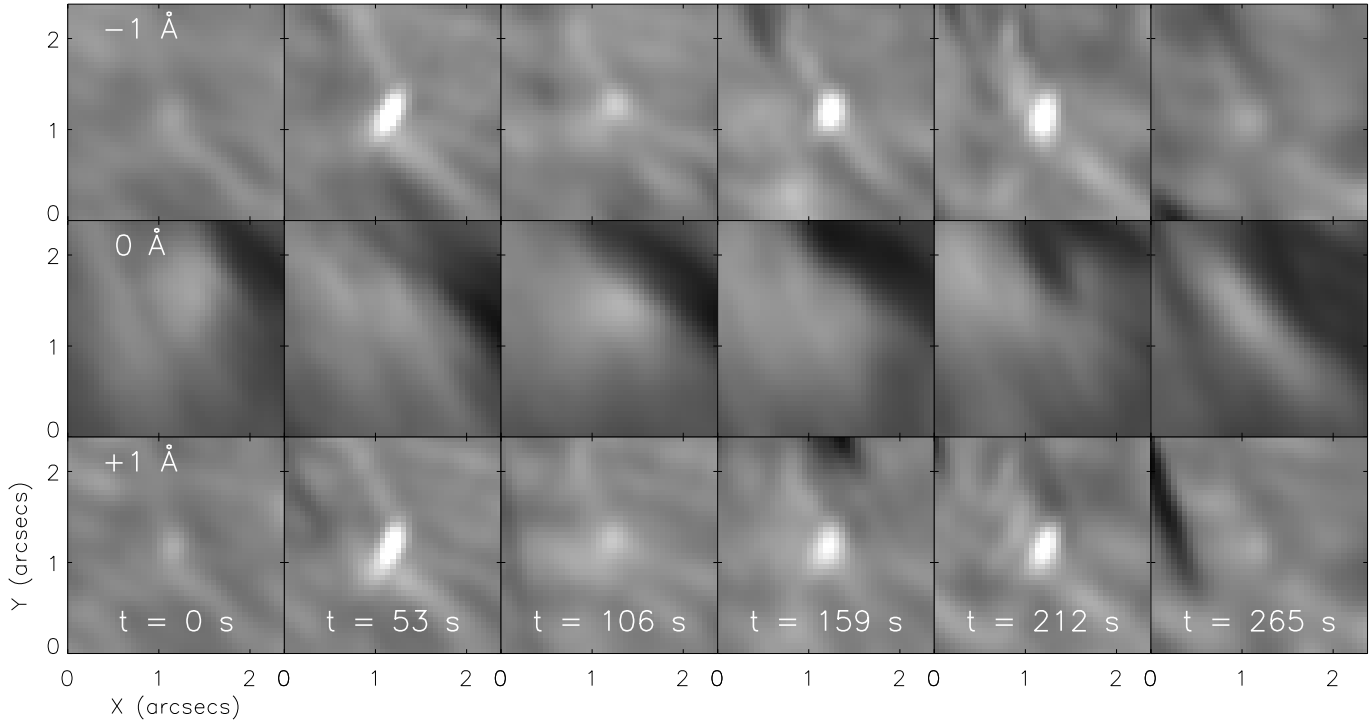}
\caption{The evolution of a repetitive QSEB over the course of around four minutes in the blue wing ($-1$ \AA; top row), line core (middle row) and red wing ($+1$ \AA; bottom row) of the H$\alpha$ line profile. A clear extension (second column) can be observed in the H$\alpha$ line wings before its retraction (third column), and re-emergence (fourth column). This behaviour is similar to the evolution of certain EBs reported in the literature.}
\label{Evol}
\end{figure*}

We begin our analysis by acquiring measurements of the lengths, widths, and lifetimes of the $21$ QSEB events studied in this article using CRISPEX. These measurements allow QSEBs to be compared with the EBs discussed in the literature (specifically by \citealt{Nelson15} who used a dataset with an identical spatial resolution thereby obtaining exactly comparable results) and to reaffirm the properties of QSEBs reported by \citet{Rouppe16}. In the top panel of Fig.~\ref{Stats}, we plot the lengths and widths of the sample of QSEBs measured here [diamonds] and the AR EBs discussed by \citet{Nelson15} [crosses]. It is immediately evident that the QSEBs are, in general, smaller than their AR counter-parts. The means of the lengths and widths of these QSEBs were found to be $0.63$\arcsec\ ($\sigma$=$0.17$\arcsec) and $0.35$\arcsec\ ($\sigma$=$0.08$\arcsec), respectively. In the bottom frame of Fig.~\ref{Stats}, we plot the length against lifetime of these QSEB features and the EBs studied by \citet{Nelson15}. The mean lifetime of these QSEBs is approximately $120$ s ($\sigma$=$60$ s), around one minute higher than the value reported by \citet{Rouppe16}. As the minimum possible lifetime of events in this sample was $53$ seconds, it is likely that this difference is due to the relatively low cadence of the data analysed here. Indeed, several events in our sample were observed to live for around $4$ minutes, with seemingly repetitive parabolic `flaming' (see Fig.~\ref{Evol}) which occurred over the course of one or two frames. For consistency, we did not classify each individual flame as a separate event, thereby increasing the mean. Recalculating the mean to account for repetition within the sample lowers the average lifetime of these QSEBs to $106$ s ($\sigma$=$47$ s).

Of the $21$ events analysed here, three were observed to display such impulsive repetitive behaviour over short time-scales during their lifetimes. This behaviour was not apparent in the majority of features studied by \citet{Rouppe16}. In Fig.~\ref{Evol}, we plot the evolution of the left-hand event presented in Fig.~\ref{Profiles} at $53$ s intervals for three positions within the H$\alpha$ line profile, namely $-1$ \AA, the line core, and $+1$ \AA. In the second column, a QSEB is easily observed which then reduces in intensity and length in the 3rd column before appearing to extend once again in the 4th column. Through a close inspection of the imaging data during this time period, it is clear that the apparent fading and contraction of the QSEB is not due to a reduction in seeing quality but is, instead, a real change in morphology of the event through time. Further to such impulsive recurrence, we also found one location where at least three seemingly independent QSEBs occured over the course of $15$ minutes. Such recurrence, shown for AR EBs (see, for example: \citealt{Qiu00, Nelson15}), is thought, in-line with the magnetic reconnection hypothesis, to be indicative of multiple releases of energy from the same spatial location, potentially due to flux build up through time (see, for example, \citealt{Reid16}). 

\begin{figure*}
\includegraphics[scale=0.4,trim={0.5cm 0 0 0}]{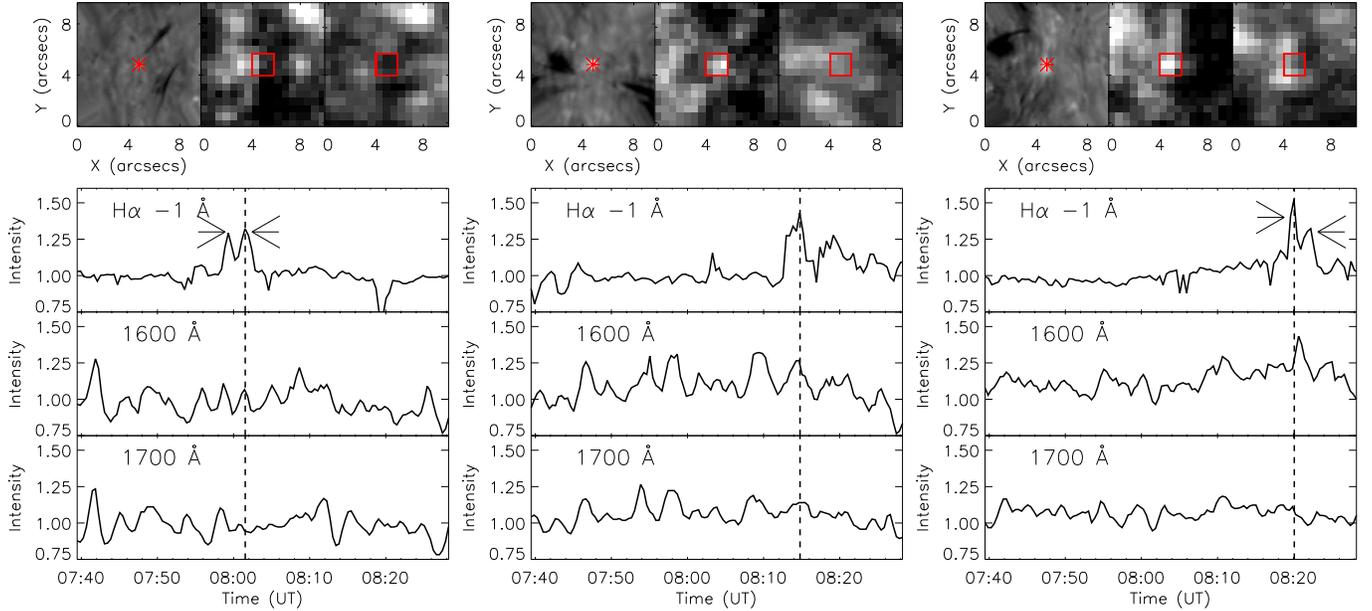}
\caption{(Top row) Context images plotting the FOV surrounding the three QSEBs plotted presented in Fig.~\ref{Profiles}. Included, respectively, from left to right for each column are: H$\alpha$ red wing ($-1$ \AA), SDO/AIA $1600$ \AA, and SDO/AIA $1700$ \AA\ images. The red crosses on the H$\alpha$ wing images and the red boxes over-laid on the SDO/AIA channels indicate the pixel/regions used to construct lightcurves. (Bottom row) Lightcurves for each QSEB made for the H$\alpha$ red wing (top panel), SDO/AIA $1600$ \AA\ channel (middle panel), and SDO/AIA $1700$ \AA\ filter (bottom panel). The dotted vertical lines indicate the time at which the QSEB reached peak intensity in the H$\alpha$ line wings, which corresponds to the frames plotted in the top row. The arrows in the left and right-hand columns highlight repetitive `flames' during the lifetimes of the QSEBs.}
\label{Lightcurve}
\end{figure*}

SDO/AIA images co-spatial to these QSEBs were also analysed. As with EBs identified in ARs (see, {\it e.g}, \citealt{Vissers13,Nelson15}), no signature of QSEBs was observed in the SDO/AIA $304$ \AA\ filter. In Fig.~\ref{Lightcurve}, we plot expanded FOVs around the three QSEBs presented in Fig.~\ref{Profiles} for the H$\alpha$ blue wing ($-1$ \AA; left-hand panel), the SDO/AIA $1600$ \AA\ filter (central panel), and the SDO/AIA $1700$ \AA\ filter (right-hand panel). The small-scale QSEBs (in the H$\alpha$ data) are almost entirely covered by the red crosses, which indicate the pixels selected to construct the lightcurves in the bottom panels. No signature was observed co-spatial to any QSEBs in the SDO/AIA $1700$ \AA\ channel, however, the events plotted in the central and right-hand columns of Fig.~\ref{Lightcurve} did appear to be linked to burst-like events in the SDO/AIA $1600$ \AA\ filter. The isolation of the UV intensity enhancement to the SDO/AIA $1600$ \AA\ data could be due to increased emission of the TR \ion{C}{4} in that filter rather than enhancements in the continuum intensity, potentially indicating that increased \ion{Si}{4} emission would also be expected. The red boxes over-laid on the SDO/AIA images indicate the regions selected to construct lightcurves.

The H$\alpha$ line wing lightcurves plotted in the top panels of the bottom row of Fig.~\ref{Lightcurve} depict the short-lived (of the order minutes) intensity increases which are indicative of the presence of the QSEBS. Both QSEBs displayed in the left (plotted through time in Fig.~\ref{Evol}) and right-hand columns are repetitive through their lifetimes, with individual peaks highlighted by the arrows. The dashed vertical lines indicate the frames plotted in the top row. The SDO/AIA $1600$ \AA\ lightcurves (middle row) plotted in the central and right-hand columns display short-lived peaks in intensity co-temporal to the formation of the QSEBs, with the right-hand column exhibiting the clearest example of this behaviour. The intensity for the minutes surrounding the QSEB approaches $150$ \%\ of the time-averaged local background intensity. These increases in SDO/AIA $1600$ \AA\ intensity, co-temporal to only a small fraction of these QSEBs, are comparable to the signatures identified co-spatial to EBs in ARs.

\subsection{Links Between QSEBs And IBs}

\begin{figure*}
\includegraphics[scale=0.54,trim={1.5cm 0 0.5cm 0}]{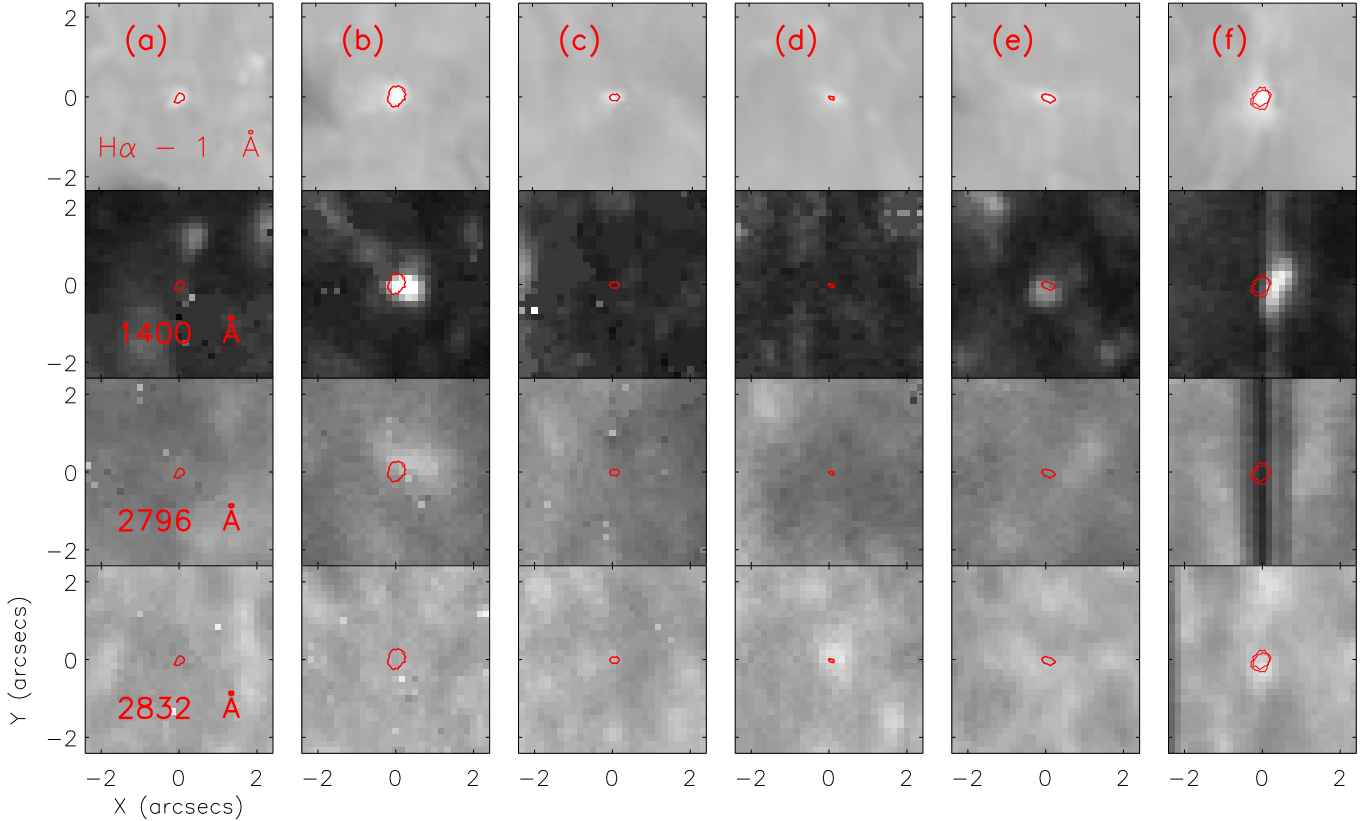}
\caption{(Top row) The six QSEBs which occurred during the transit of the IRIS SJI across the SST/CRISP FOV. The features in the second and sixth columns have previously been presented in Fig.~\ref{Profiles} and Fig.~\ref{Lightcurve} (in the second and third columns of these figures, respectively). (Second row) The IRIS SJI $1400$ \AA\ response to these events, depicting the co-spatial formation of IBs (in the second and sixth columns). The co-spatial IRIS SJI $2796$ \AA\ (third row) and $2832$ \AA\ (bottom row) data are also included for completeness. The red contours indicate the pixels over $130$ \% of the background intensity in the H$\alpha$ line wings ({\it i.e.}, the locations of the QSEBs).}
\label{IRISSJI}
\end{figure*}

As IRIS was conducting a raster from off the limb to the solar disk, only the six QSEB features which occurred between $08$:$10$:$00$ UT and $08$:$20$:$00$ UT were studied using data collected by the SJI. This sample included the features plotted in the central and right-hand columns of Fig.~\ref{Profiles} and Fig.~\ref{Lightcurve}. In Fig.~\ref{IRISSJI} we plot the $1400$ \AA\ (second row), $2796$ \AA\ (third row), and $2832$ \AA\ (bottom row) responses to the QSEBs (plotted in the H$\alpha$ blue wing in the top row). Of these events, two (the QSEBs displayed in the second and sixth columns which correspond to the central and right-hand events of Fig.~\ref{Lightcurve}, respectively) were identified to form co-spatial to short-lived increases in intensity in the \ion{Si}{4} $1400$ \AA\ filter, which appeared analogous to IBs (see, for example, \citealt{Peter14,Vissers15,Tian16}). None of the other four events were observed co-spatial to IBs, agreeing with the results of \citet{Rouppe16} that the majority of QSEBs display no TR signature. The two QSEBs which formed co-spatial to obvious IRIS signatures were neither particularly large, intense, or long-lived, appearing to be similar to the majority of QSEBs in this sample.

\begin{figure*}
\includegraphics[scale=0.44,trim={0.6cm 0 0 0}]{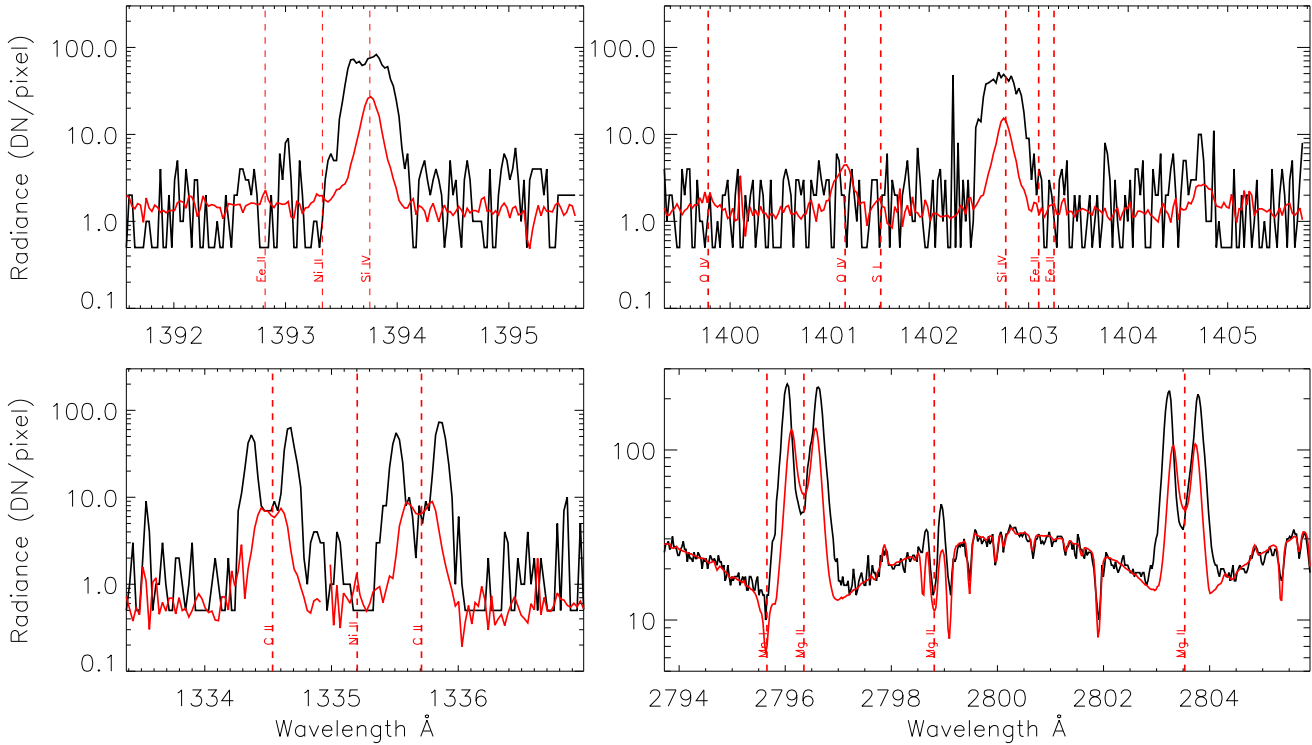}
\caption{(Top row) The left and right columns, respectively, plot \ion{Si}{4} $1394$ \AA\ and \ion{Si}{4} $1403$ \AA\ spectral profiles for the IB pixel (black line) and an averaged reference profile (red line). (Bottom row) Spectral profiles for the \ion{C}{2} and \ion{Mg}{2} doublets sampled at the same locations as the top row. Note the logarithmic scaling of the $y$-axis. Vertical red dashed lines (and the corresponding labels) indicate spectral locations of specific interest to this analysis.}
\label{IRISspec}
\end{figure*}

Interestingly, the feature in the right-hand column of Fig.~\ref{IRISSJI} was also sampled by the IRIS slit during its lifetime displaying an IRIS burst-like spectrum. In the top row of Fig.~\ref{IRISspec}, we plot the \ion{Si}{4} $1393$ \AA\ (left) and \ion{Si}{4} $1403$ \AA\ (right) spectral profiles sampled at the location of the IB (black line) and averaged over a quieter region close to this feature (red line). The bottom row plots the \ion{C}{2} (left) and \ion{Mg}{2} (right) spectra. Wavelength calibration was conducted following the method suggested by \citet{Tian16}. The \ion{Si}{4} $1393.755$ \AA\ window Doppler shift was estimated using the \ion{Ni}{2} $1393.330$ \AA\ line (which was assumed to have zero Doppler shift). This shift was then also applied to the \ion{Si}{4} $1402.770$ \AA\ window. The \ion{C}{2} spectral window was calibrated using the \ion{Ni}{2} $1335.203$ \AA\ line (which, again, was assumed to have no Doppler shift). The accuracy of the detected shifts were confirmed by the similarity of the shifts in the \ion{Ni}{2} $1393.330$ \AA\ and \ion{Ni}{2} $1335.203$ \AA\ lines. The shift in the \ion{Mg}{2} window was estimated using the neutral \ion{Mn}{1} $2795.633$ \AA\ line. 

The general shapes of the plotted spectral lines sampled at the location of the QSEB are similar to IBs discussed in the literature (see, for example, \citealt{Peter14,Tian16} for a variety of IB spectra), including wider and brighter \ion{Si}{4} profiles, increases in the \ion{Mg}{2} and \ion{C}{2} line wing intensities, and absorption profiles of some chromospheric lines (for example, \ion{Ni}{2} $1335.203$ \AA, \ion{Ni}{2} $1393.330$ \AA, \ion{Fe}{2} $1403.225$ \AA). The ratio between the intensities of the \ion{Si}{4} $1393$ \AA\ and \ion{Si}{4} $1403$ \AA\ lines at the location of the QSEB is $\sim1.71$, lower than that calculated for the reference profiles of $\sim1.80$ and the optically thin case of $2$. In addition to this, some absorption is observed in the core of the \ion{Si}{4} $1393$ line, consistent with the self absorption discussed by \citet{Yan15}. These profiles provide the first evidence that certain QSEBs can occur co-spatial to IBs.

Quantitatively, the increases in intensity and line-width measured co-spatial to this QSEB are around an order of magnitude smaller than those previously reported around IBs (although it should be noted they are similar to some examples, including IB $5$ discussed by \citealt{Tian16}). This could be expected, however, given the reduced sizes and lower H$\alpha$ line wing intensities of QSEBs in comparison to AR EBs (as well as the typically lower background intensities in the quiet-Sun). Interestingly, the \ion{Si}{4} line widths measured for this IB are smaller than observed for most IBs in the literature. Potentially, this could be caused by the viewing angle as, if the bi-directional jets associated with the QSEB are predominantly vertical, only the limited line-of-sight component of such motions would be measured; however, this is currently only speculation.
	
\begin{figure*}
\includegraphics[scale=0.4,trim={-2cm 0 0 0}]{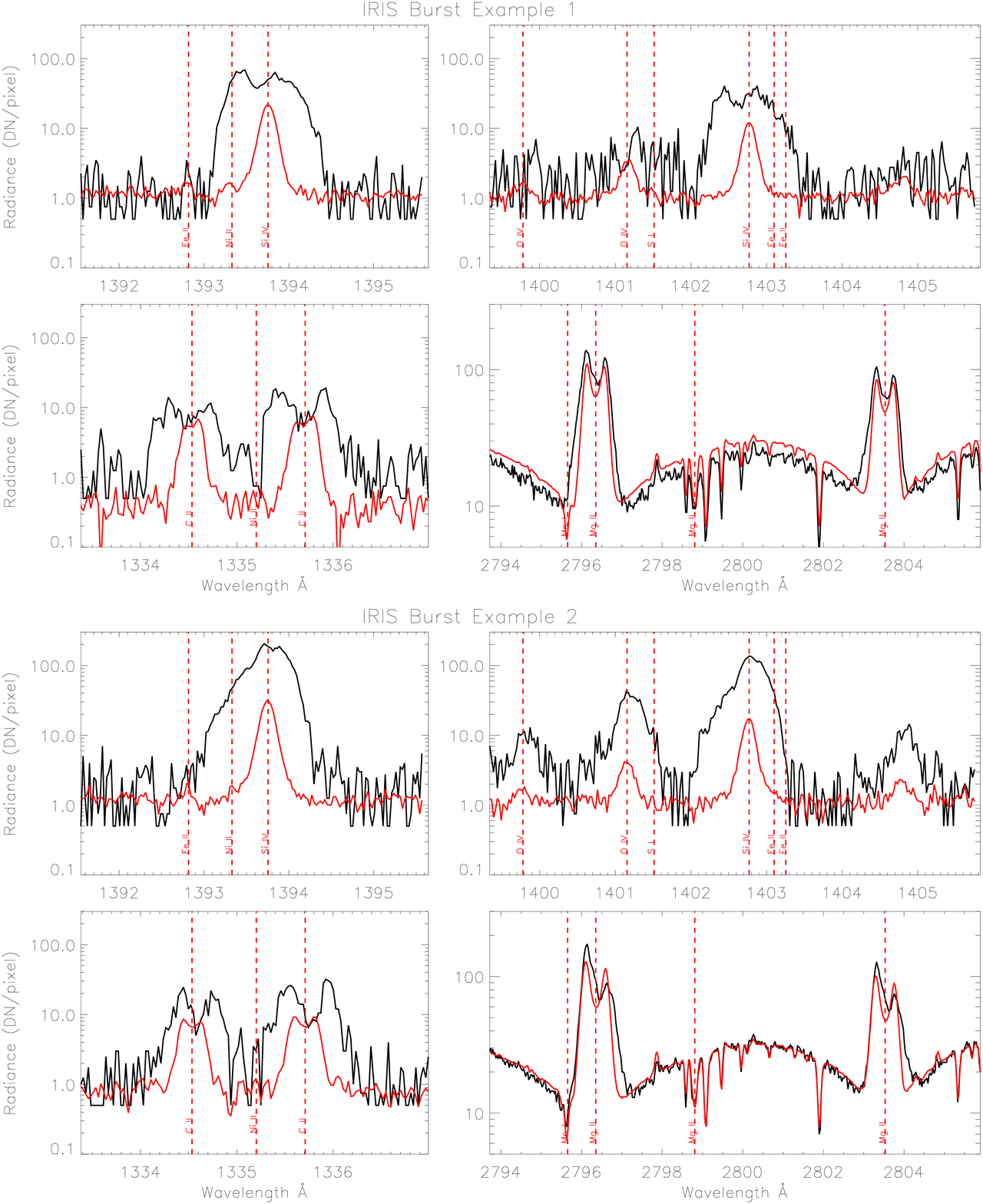}
\caption{Same as Fig.~\ref{IRISspec} but for two further examples of IRIS burst-like events observed in these data.}
\label{nonIRISspec}
\end{figure*}

In Fig.~\ref{nonIRISspec}, we plot spectral data for two further IBs (with the same layout as applied in Fig.~\ref{IRISspec}) which occurred after the SST/CRISP instrument had stopped acquisition. Therefore, these events could not be linked to any QSEB. The \ion{Si}{4} spectra for both of the events plotted in Fig.~\ref{nonIRISspec} displayed much larger line widths than the event plotted in Fig.~\ref{IRISspec} appearing to be analogous to the features discussed by \citet{Tian16}. The ratios between the two \ion{Si}{4} lines are $1.83$ and $1.47$ for Examples $1$ and $2$, respectively, and again indicate a departure from the optically thin regime. It should be noted that the intensity enhancement in the \ion{Si}{4} line cores for Example $2$ are twice those measured for the QSEB-linked IB and are, therefore, comparable to the intensities measured co-spatial to IBs in ARs by \citet{Tian16}.

The \ion{C}{2} data for both IBs display broadened and enhanced line wings. The \ion{Mg}{2} h\&k peak intensities in Fig.\ref{nonIRISspec} are also slightly asymmetric, most likely due to velocity gradients in the atmosphere shifting the wavelength of maximum opacity to the red causing increased emission in the blue peak. This effect has been described in detail in \citet{Carlsson97} and observed in flare line profiles presented in \citet{Kuridze15} (for H$\alpha$) and \citet{Kerr16} (for \ion{Mg}{2}). Both of the events plotted in Fig.~\ref{nonIRISspec} also show evidence of self-absorption in the \ion{Si}{4} $1393$ \AA\ line, with Example $1$ displaying clear self-absorption in the \ion{Si}{4} $1403$ \AA\ line as well. This self-absorption is also consistent with the scenario suggested by \citet{Yan15}, whereby the increased density within the feature causes absorption at the line core.

\subsection{H$\alpha$ and \ion{Ca}{2} $8542$ \AA\ line synthesis}

Finally, we investigate one-dimensional RADYN simulations (\citealt{Carlsson92, Carlsson95}) created by perturbing three quiet solar-like atmospheres by depositing energy at a range of heights (building on the work recently presented by \citealt{Reid17}). The motivation of this work is to attempt to explain the profiles displayed in Fig.~\ref{Profiles}, where the H$\alpha$ wing intensities are enhanced but no \ion{Ca}{2} $8542$ \AA\ response is observed. \citet{Rouppe16} suggested that this observational signature is caused by the occurrence of the QSEB at heights which do not influence the \ion{Ca}{2} $8542$ \AA\ line. Whether such layers exist still requires verification. We conducted a large range of simulations where either $100$ erg cm$^{-3}$ s$^{-1}$, $300$ erg cm$^{-3}$ s$^{-1}$, or $500$ erg cm$^{-3}$ s$^{-1}$ of energy was inputted into a static atmospheric model at deposition layers (ranging from the photosphere to the chromosphere) which had heights of either $50$ km or $200$ km. We allowed the systems to stablise for $9$ s, before H$\alpha$ and \ion{Ca}{2} $8542$ \AA\ line profiles were constructed after $10$ s of solar time using the MULTI package built into RADYN. Overall, more than $70$ models were considered.

Before considering the results of these simulations, we briefly discuss the reasons for selecting the parameters introduced in the previous paragraph. Firstly, we considered energy deposition layers of $50$ km and $200$ km as smaller values would be below the spatial resolution of instruments such as the SST/CRISP and larger layers would be well above the local scale height. Any energy deposition layers larger than $200$ km would, therefore, not satisfy the condition that energy is deposited at a preferential location in the solar atmosphere, causing the observed spectral profiles of QSEBs. Secondly, the $500$ erg cm$^{-3}$ s$^{-1}$ energy deposition rate was found to produce unrealistically high intensity enhancement in the H$\alpha$ line profiles. Therefore, we do not consider any energy deposition rates higher than this. Energy deposition rates lower than $100$ erg cm$^{-3}$ s$^{-1}$ did not provide the required H$\alpha$ wing intensity increases and were also not considered. Finally, three different starting atmospheres (one quiet Sun, as well as QS.SL.LT and QS.SL.HT from \citealt{Allred15}) were studied although the results obtained for each were comparable and, as such, we only present results from the quiet Sun atmosphere.

\begin{figure*}
\includegraphics[scale=0.59]{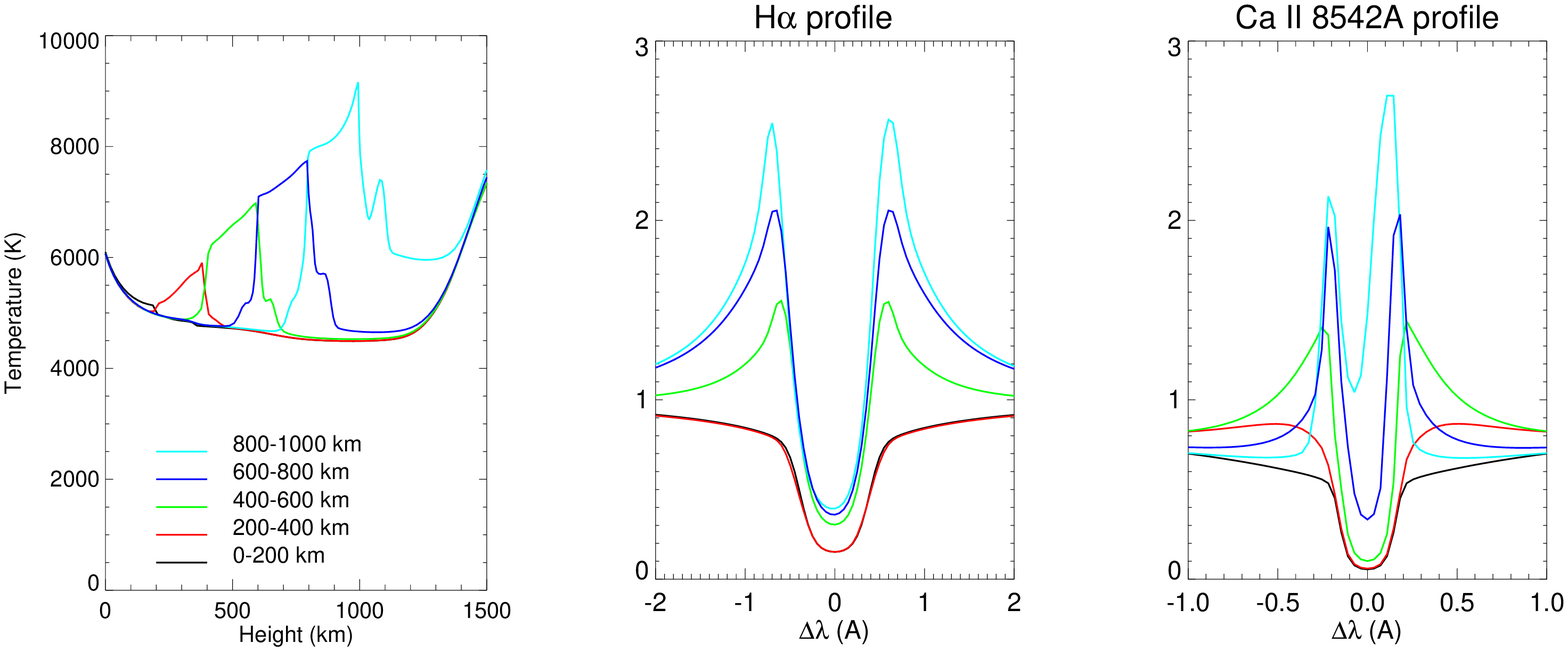}
\caption{Five temperature profiles (left panel) corresponding to five $200$ km high energy deposition set-ups. The heights of each coloured line are indicated in the legend. The H$\alpha$ (centre panel) and \ion{Ca}{2} $8542$ \AA\ line profiles synthesised for each one-dimensional atmosphere.}
\label{Synth_Profiles}
\end{figure*}

In the left hand panel of Fig.~\ref{Synth_Profiles}, we plot examples of temperature profiles for five $200$ km high energy  ($300$ erg cm$^{-3}$ s$^{-1}$) deposition layer simulation runs, measured at $t=10$ s. Each coloured line in the left-hand panel denotes a different energy deposition layer. These examples are representative of the entire suite of models which we studied. The centre and right hand panels plot the respective H$\alpha$ and \ion{Ca}{2} $8542$ \AA\ synthesised line profiles. It is immediately evident that the higher the energy is deposited in the atmosphere, the more emission the synthesised wings of both the H$\alpha$ and \ion{Ca}{2} $8542$ \AA\ lines display. In no case do the synthesised H$\alpha$ wings form in emission when the \ion{Ca}{2} $8542$ \AA\ wings do not. On the contrary, when the energy is deposited between $200$-$400$ km, enhanced \ion{Ca}{2} $8542$ \AA\ line wings are evident with no response from H$\alpha$. Qualitatively, these results do not change when one considers the shorter $50$ km energy deposition bins. The line core intensities of both lines are also increased for higher energy deposition layers, although, this is probably due to the lack of three-dimensional effects ({\it i.e.}, the lack of overlying canopy) in the simulations. Varying the energy deposition rate only changes the level of enhancement across both lines.

\begin{figure}
\includegraphics[scale=0.54,trim={-1cm 0 3cm 0.5cm}]{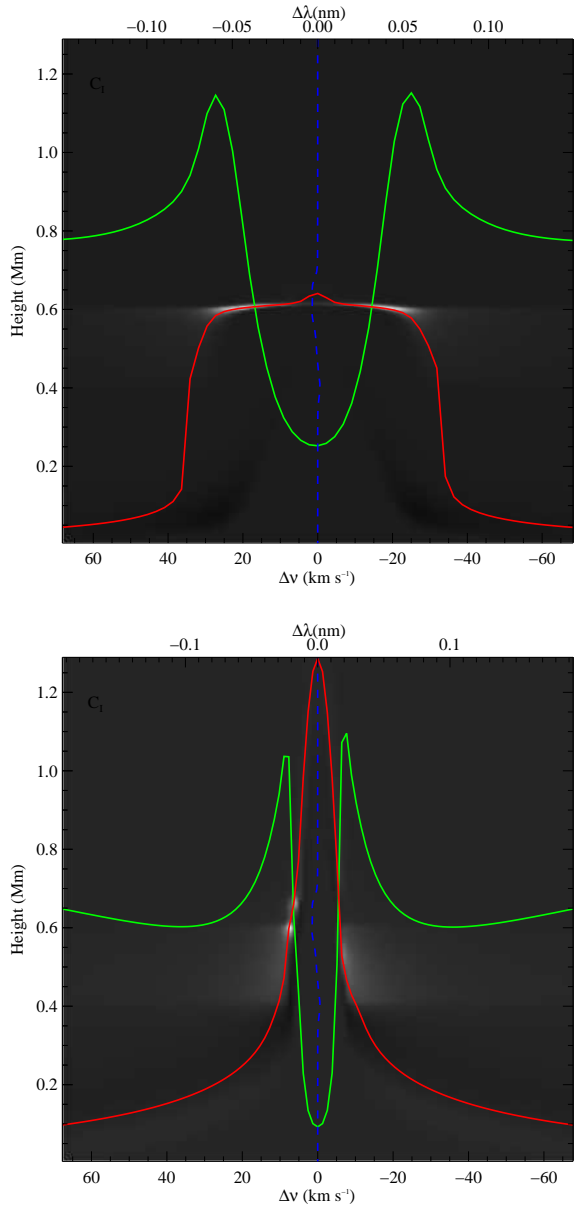}
\caption{(Top panel) The difference between the contribution functions at $t=0$ s and $t=10$ s for the H$\alpha$ line profile. The line profile at $t=10$ s is overlaid in green. The red line indicates the $\tau=1$ level. (Bottom panel) Same as above but for the \ion{Ca}{2} $8542$ \AA\ line profile.}
\label{Cont_func}
\end{figure}

The strong connection between H$\alpha$ line wing increases and \ion{Ca}{2} $8542$ \AA\ line wing increases in these simulations comes from the modification of the contribution function of both lines due to the energy deposition. In the top panel of Fig.~\ref{Cont_func}, we plot the difference between the $t=0$ s and $t=10$ s contribution functions for the H$\alpha$ line profile, for a representative $200$ km high energy deposition layer. The overlaid lines plot the synthesised line profile (green; arbitrary scaling) and the $\tau=1$ height (red). The bright regions at a simulation height of $400$ km and a Doppler shift of $\pm20$ km s$^{-1}$ indicate the locations at which the enhanced emission in the H$\alpha$ line wings (typical of EBs and QSEBs) occurs. The bottom panel plots the corresponding information for the \ion{Ca}{2} $8542$ \AA\ line profile. This differenced contribution function also displays the bright regions (which are, perhaps, even more obvious than in H$\alpha$) at Doppler shifts of around $\pm20$ km s$^{-1}$, which lead to increases in intensity in the \ion{Ca}{2} $8542$ \AA\ line wings. Due to the change in the source function and opacity caused by the heating, the contribution of both the H$\alpha$ and \ion{Ca}{2} $8542$ \AA\ line wings increases at these heights, regardless of the quiet-Sun formation heights of these lines. Overall, these results do not appear to support the assertion that energy deposition, consistent with EBs and QSEBs, at specific heights in the solar atmosphere can lead to H$\alpha$ wing emission without \ion{Ca}{2} $8542$ \AA\ wing intensity enhancements. However, future work should aim to investigate this further, perhaps using different starting atmosphere models.

\section{Discussion and Conclusions}
	\label{Discussion}

QSEBs are an interesting newly discovered phenomenon, identified by increases in intensity in the H$\alpha$ line wings similar to EBs (\citealt{Ellerman17,Nelson15,Vissers15}) but located in the quiet-Sun. These events are thought to highlight the occurrence of magnetic reconnection in the photosphere outside of ARs, perhaps similar to the modelling presented by \citet{Nelson13} and \citet{Danilovic17}. In this paper, we have not only corroborated the results of \citet{Rouppe16}, but have also highlighted some new properties of these features, including repetition over both short and long time-scales (in comparison to the lifetimes of the QSEBs), and shown the presence of an IB co-spatial to an individual QSEB (similar in nature to IB occurrence co-spatial to EBs; \citealt{Tian16}), thereby also confirming the presence of IBs in the quiet-Sun. In the following paragraphs we shall present a brief overview of our results and discuss how they fit in with the current understanding of small-scale reconnection events.

Initially, $21$ QSEBs were identified in H$\alpha$ line scans collected at the solar limb by the SST/CRISP instrument. The basic properties of these events were comparable to those found by \citet{Rouppe16} with average lifetimes, lengths, and widths of approximately $120$ s, $0.63$\arcsec, and $0.35$\arcsec, respectively. These values are at the lower end of the spectrum of properties previously derived for EBs in ARs, as is shown in Fig.~\ref{Stats}. Two features within this sample did appear co-spatial to limited ($<120$ \%) \ion{Ca}{2} $8542$ \AA\ wing brightenings, however, as the bright regions in the \ion{Ca}{2} $8542$ \AA\ line wings bore little resemblence to the clear, elongated QSEBs identified in H$\alpha$, it is likely that the \ion{Ca}{2} $8542$ \AA\ heightened wing emission was not related to the QSEB. By studying a large range of RADYN simulated profile, created by perturbed reference profiles by an input of energy, we were unable to reproduce line profiles which displayed enhanced H$\alpha$ line wing emission and no \ion{Ca}{2} $8542$ \AA\ response (see Fig.~\ref{Synth_Profiles}). 

Repetitive, impulsive flame-like behaviour (shown to be common for EBs; see, {\it e.g.}, \citealt{Vissers15,Nelson15}) was observed for three QSEB events in our sample. The evolution of one of these QSEBs is detailed in Fig.~\ref{Evol} and the lightcurves constructed for that event and one further example are plotted in the left and right-hand columns of Fig.~\ref{Lightcurve}, respectively, with the repetitive peaks indicated by the arrows. Such repetition was not widely observed either here or by \citet{Rouppe16}, begging the question as to whether recurrence is common across the solar disk or whether it is only limited to certain regions with currently unknown similarities where, perhaps, flux build-up occurs more readily (for example, at super-granular boundaries). 

Two of these QSEBs formed co-spatial to burst events in the SDO/AIA $1600$ \AA\ UV and SJI $1400$ \AA\ data, with one of these events being sampled by the IRIS slit during its lifetime. The IRIS spectra displaying increased intensity in the \ion{Si}{4} $1393$ \AA\ and $1403$ \AA\ lines, as well as wing intensity increases in the \ion{C}{2} and \ion{Mg}{2} spectral windows (see Fig.~\ref{IRISspec}). These profiles were analogous to IBs discussed in the literature (\citealt{Peter14, Vissers15, Tian16}). The \ion{Si}{4} line widths were smaller than the majority of IBs, however, this could be due to line-of-sight effects if the dominant motion of the QSEB is vertical away from the solar disk ({\it i.e.}, perpendicular to the line-of-sight). Support for this assertion was found through analysis of several other IBs, which were identified in these data (presented in Fig.~\ref{nonIRISspec}). These IBs displayed larger \ion{Si}{4} line widths in addition to blue shifted \ion{Mg}{2} profiles potentially indicating velocities in the line-of-sight.

Overall, our results indicate that the majority of QSEBs are smaller and apparently weaker than their AR cousins, agreeing with the results of \citet{Rouppe16}. However, the IBs co-spatial to two of these features indicate that that some (likely a small minority of) QSEBs could be linked to localised heating of plasma to TR temperatures, in a similar manner to energetic EBs in ARs. Future observational work should be carried out to discover how common such apparently energetic QSEBs are. It also remains to be seen whether the interesting H$\alpha$ and \ion{Ca}{2} $8542$ \AA\ signatures of QSEBs can be reproduced through further semi-empirical modelling.

\acknowledgements
IRIS is a NASA small explorer mission developed and operated by LMSAL with mission operations executed at NASA Ames Research center and major contributions to downlink communications funded by ESA and the Norwegian Space Centre. The Swedish $1$-m Solar Telescope is operated on the island of La Palma by the Institute for Solar Physics of Stockholm University in the Spanish Observatorio del Roque de los Muchachos of the Instituto de Astrof\'isica de Canaries. This work was inspired by discussion at the `Solar UV bursts -- a new insight to magnetic reconnection' meeting at the International Space Science Institute (ISSI) in Bern. We would like to thank the anonymous referee for useful comments which improved this work. CJN and RE thank the Science and Technology Facilities Council (STFC) for the support received to conduct this research. RE acknowledges the support received by the CAS Presidents International Fellowship Initiative, Grant No. 2016VMA045 and is also grateful to the Royal Society for their support. RO acknowledges support from MINECO and FEDER funds through grant AYA2014-54485-P. We also thank Luc Rouppe van der Voort for invaluable help with data reductions.

\bibliographystyle{apj}
\nocite{*}
\bibliography{QSEBs}

\end{document}